\documentclass[a4paper,10pt,twocolumn]{article}
\onecolumn
\title{Reinterpreting the Pioneer anomaly and its annual residual}
\author{Paul G. ten Boom\footnote{email address: \textsf{ptb@phys.unsw.edu.au}}\\\em{School of Physics, University of New South Wales, Sydney, N.S.W. 2031, Australia}}
\begin{document}
\maketitle
\begin{abstract}
In addition to its long-term constancy, the Pioneer (spacecraft)
anomaly appears to only exist for bodies whose mass is less than
that of: planets, moons, comets, and heavy asteroids of known
mass. Assuming the observational evidence is reliable and not the
result of an unknown systematic effect, a violation of the Weak
Principle of Equivalence is implied. This constraint is the most
confronting for any prospective new physics. Any new physical
mechanism that proposes an additional gravitational force, i.e. an
additional spacetime curvature, is rendered unreasonable because
all masses should be equally affected. This paper examines an
approach that is based upon the existence of a sum of additional
field energies. A finite number of tiny wavelike undulations, upon
the existing gravitational field, are hypothesized. The sources of
these non-Einsteinian gravitational waves are the spin-orbit
coupled moons of the solar system. An excess energy arising from:
a lunar orbital motion, quantum mechanical geometric phase and
spacetime curvature generates these new gravitational field waves.
General Relativity's Lorentz invariance demands that these
``acceleration-waves" have constant amplitude. The dissipation of
these spherical waves, as they expand out from suitable moons, is
seen to exist as a volumetric based reduction in the inertial mass
that can sympathetically oscillate with these waves. Therefore,
masses above a given wave's cut-off mass remain completely
unaffected. The full substantiation and quantification of these
undulations upon the gravitational field is deferred to another
paper. This article simply seeks to show that a superposition of
these ``acceleration-waves" results in an oscillatory or
non-steady expression of spacecraft kinetic energy. The
superposition of these additional oscillatory components of
longitudinal motion leads to a shortfall of actual steady
translational motion, as compared to predictions that assume a
purely steady motion. With wave energy proportional to lunar mass,
and some moons inactive for geometric reasons, the four Galilean
moons of Jupiter and Saturn's Titan dominate this new effect. The
orbital resonances of Jupiter's Galilean moons, markedly
attenuates the variation in magnitude of the Fourier-like
superposition of these waves. The small variance observed for the
Pioneer (acceleration) anomaly around its long-term average has
created the misleading impression of a constant additional inward
acceleration with associated observational noise. Additionally,
Jupiter's least orbitally resonant moon Callisto, and Saturn's
Titan, cause the Pioneer spacecraft to exhibit a (synodic) 356 day
resonance, that has been misinterpreted as an Earth based `annual'
residual. The amplitudes of the cyclic diurnal and annual
Earth-to-spacecraft motion `offsets' are examined in some detail.
The Earth based annual residual is found to be actually very small
and incapable of figuring in the Doppler tracking observations. In
order to clearly establish and scrutinize the physical validity of
the hypothesis being presented, aspects of the author's broader
model are incorporated into the discussion only where necessary.
Direction cosine corrections, for wavefront direction relative to
spacecraft trajectory, are quantitatively neglected in order to
primarily scrutinize the physical viability of the hypothesis.
Finally, the hypothesis sheds promising light upon other concerns
regarding the modelling of gravitation influencing `light' bodies
in our solar system. These include: the ``Earth Flyby Anomaly", an
apparent absence of small comets, an apparent paucity of smaller
bodies in the Main Belt of asteroids, and residual doubts
concerning the ``Migrating Planets" hypothesis - that addresses
the too rapid formation of the ice giants Uranus and Neptune.
\end{abstract}
\twocolumn
%-------------------------------------------------------------------------------------------------------------
\section{Introduction}
The Pioneer anomaly has failed to be explained by any systematic
effects and has two awkward observational characteristics. The
first is its (long term) constancy and the second is that it
implies an apparent violation of the principle of equivalence ---
in that only spacecraft and not: planets, moons, comets or large
asteroids of known mass, seem to be affected~\cite[p.3]{Bert04}.
By way of these two awkward observational constraints the anomaly
has resisted full explication. The latter constraint and
observational evidence appear to preclude any standard force based
hypothesis, such as a modification of Einstein's General
Relativity. General Relativity, as a theory of gravitation, holds
impressively in the solar system for electromagnetic wave
propagation and the motions of `heavy' bodies.

A hypothesis concerning the re-expression of a tiny proportion of
total spacecraft kinetic energy into a number of (longitudinal)
oscillatory components of kinetic energy is proposed. These
coexist with and in addition to the dominant steady
(non-oscillatory) kinetic energy. These additional `spectral'
components, together, then cause an ongoing shortfall in actual
speed relative to predicted spacecraft translational speed. This
motion is a response to tiny harmonic\footnote{In order to avoid
confusion with wave resonances, the expression `harmonic' is not
used to describe sinusoidal \em{and} \rm co-sinusoidal functions.
The term `sinusoidal' is preferred.} undulations of the
gravitational field.
\subsection{Outlining a new approach} This write up primarily
presents the ideas and mathematics that illustrate how the
anomalous Pioneer `acceleration' can alternatively be seen as
potentially a Fourier summation of first order constant amplitude
(acceleration or gravitational\footnote{To avoid confusion with
Einstein's gravitational waves, the expression `acceleration
waves' is preferred. These acceleration waves, like gravitation,
produce spacetime curvature.}) fluctuations upon the pre-existing
gravitational field. To support this it is argued that, unlike the
Earth based diurnal residual, the `annual' residual is spacecraft
based and hence real (and not quite of 365.25 days duration
either). This residual is modelled to be the only obvious
resonance of  a number of sinusoidal undulations on (and of) the
gravitational field. The amplitudes of these sinusoidal
undulations are very small compared to the strength of the
gravitational field. Only in very weak gravitational fields is
their presence significant.

The source of these waves shall not be fully
elucidated\footnote{The physical establishment of the mechanism is
vital, but this involves a long write up that is primarily quantum
mechanical in nature. The reader is asked to `bear with the
hypothesis'.}, other than to say, that observational evidence and
the author's model dictate that they originate from regular moons
of the solar system held in spin-orbit coupling around their
respective host planets (which in turn orbit the Sun)\footnote{A
fully detailed model produces amplitudes of the waves, and the
distribution of the cut-off mass with respect to distance from the
(finite number of) sources.}.

This new mechanism is dominated by the bigger moons of the solar
system. Of the big seven moons --- Earth's moon (Moon, or
unofficially Luna) and Neptune's Triton do not `generate'
acceleration waves upon the pre-existing gravitational field.
Thus, Jupiter's four Galilean moons and Saturn's Titan are seen to
dominate the mechanism's effect in our solar system\footnote{The
collision based origin of the Earth's moon, with its large angular
momentum, and the retrograde motion of Neptune's Triton makes
their relationship between third-body orbital kinematics,
spacetime curvature, quantum mechanical indeterminacy, and
geometric phase advance incapable of generating any atomic virtual
`excess' energy. Since all atoms of some moons collectively share
the same \em{virtual} \rm excess spin angular momentum based
energy these virtual energies `sum' to become a singular \em{real}
\rm excess energy that is expressed upon the gravitational field
as a new kind of \em{wave}\rm.}.
%------------------------------------------------------------------------------------------------------------------------
\section{Expanding the hypothesis}
In this section the basic preliminary material required to support
both: a new model, and the mathematics that follows, is presented.
\subsection{A Fourier wave summation}
The constancy of the Pioneer anomaly means that any Fourier-like
summation of first order fluctuations of acceleration upon the
gravitational field would need to be \em{special} \rm in that:
they are of constant amplitude, and together they display only a
hint of abnormally large variance around a constant
mean~\cite[Fig.\,14, p.24]{And02}.

With three of Jupiter's four Galilean moons in orbital resonance,
and Ganymede and Callisto in a subtle 7 to 3 resonance; only Titan
is a free `agent'. These orbital resonances act to smooth out the
variations in the Fourier summation of these acceleration (or
gravitational field) waves. Thus a superposition of waves
approaches a constant amplitude but this is never attained in the
short term.
\subsection{(Acceleration) Wave aspects}
This section is well short of fully comprehensive. An
\em{asymmetrical} \rm interaction of: gravitational curvature of
space, the quantum mechanical geometric phase of atoms, and
quantum indeterminacy of energy (per orbit) in the third body of a
three body celestial system leads to these `acceleration' waves
upon (and of) the gravitational field. Indeed they violate the
usual scope of energy conservation, although, it should be noted
that conservation laws rely upon symmetry.
\begin{quote}
... where there is a symmetry there is a conservation law, and
with certain reservations the converse is also
true~\cite[p.159]{Park92}.
\end{quote}
Geometric phase advance in prograde celestial three body motion,
involving only positive mass, is an inherently asymmetrical
situation\footnote{A force based model of the union of quantum
mechanics and general relativity always conserves energy, \em{a
priori}\rm. By way of an overlooked energy (imbalance and
subsequent) transfer, e.g. from the indeterminate micro
(`sub'-quantum) world to the macro (specetime curvature) world,
this force based approach could be found wanting.}.

General Relativity's Lorentz invariance demands constant
amplitudes. The dissipation of the waves is enacted by a wave
volume based reduction in the mass that these waves can influence
to sympathetically `oscillate' in response to them. The waves
have: an overall spherical shape, a period matching the
heliocentric period of orbiting moons, the propagation (of
constant wave phase) is at the speed of light, and the wave's
particles are necessarily spin~0 or spin zero\footnote{Thus, the
waves carry no angular momentum. The waves only `carry' energy and
yet their origin removes the \em{excess} \rm virtual angular
momentum per orbit shared by numerous atoms of certain moons. When
such waves are generated, the momentum of a moon, as a bulk
object, remains unchanged.}. The particles associated with these
waves impart no momentum to the Pioneer spacecraft (or any other
body.) Thus, it is only the undulations upon the gravitational
field that are capable of physically influencing (`lighter')
bodies in the solar system.
\subsection{Regarding the equivalence principle}~\label{Sect:Eq}
Regarding the equivalence principle Thibault Damour offers the
following advice.
\begin{quote} The Equivalence Principle (EP) is a
heuristic hypothesis which was introduced by Einstein in 1907, and
used by him to construct his theory of General Relativity. [The]
EP is not one of the basic principles of Nature (like, say, the
Action Principle, or the correlated Principle of Conservation of
Energy). It is a ``regional" principle, which restricts the
description of one particular interaction [mediated by a massless
spin-2 field]. An experimental ``violation" of the EP would not at
all shake the foundations of physics (nor would it mean that
Einstein's theory is basically ``wrong"). Such a violation might
simply mean that the gravitational interaction is more complex
than previously assumed, and contains, in addition to the basic
Einsteinian spin-2 interaction, the effect of another long-range
field. (From this point of view, Einstein's theory would simply
appear as being incomplete.)~\cite{Thib01}
\end{quote}
Einstein's Principle of Equivalence is (non-locally) restricted to
uniform fields, and thus it does not conflict with undulations
placed upon a (predominantly static) gravitational field that are
not solely of gravitational origin\footnote{In the broadest sense,
it is an \em{interaction} \rm between general relativity and
quantum mechanical (energy) indeterminacy (over a `cycle' time)
that is involved in the establishment of the undulations upon the
gravitational field. An alternative name for these waves is
``gravito-quantum radiation".}. Indeed, a uniform gravitational
field may be seen as an oscillatory field whose frequency goes to
zero, and whose period goes to infinity.

The new model being proposed, in line with the observational
evidence, violates the weak principle of equivalence in that the
existence or non-existence of a body's oscillatory acceleration
response to the field waves is (inertial) mass
dependent\footnote{Additionally, the wave's energy is dependent
upon the \em{number} \rm of atoms or molecules in a moon and hence
the \em{composition} \rm of a moon shall also have a slight effect
upon the total amount of wave energy a moon (of given mass) may
`release' or `generate'.}.
\subsection{Gravitation \& potential energy} \label{Sect:GandE}
In General Relativity, potential energy (P.E.) as a sum of
particle energies is not well defined, because potential energy
lies in the geometry of spacetime itself. Additionally, we may say
gravitational energy cannot be localized. The following quote from
Michael Mensky reinforces this point of view.
\begin{quote}The question of conserving energy-momentum in General Relativity
(GR) always attracted much attention. One of the reasons is that
covariant description of energy-momentum seems to be incompatible
with the integral conservation law. Particularly, it is generally
believed that no integral conservation law follows from the
covariant differential conservation law for the energy-momentum
tensor (EMT) of matter (i.e., from its covariant divergence being
zero)~\cite[p.261]{Mensky04}.
\end{quote}
Today we may still think in terms of P.E. but attempting to
quantitatively relate this to GR appears to be very awkward, if
not ill-conceived. On the contrary, relating P.E. to
classical-like waves or `ripples' of acceleration, i.e. spacetime
curvature undulations, is conceptually simple. The idea that a
wave contains energy is familiar; thus an acceleration wave can be
seen to possess some sort of potential energy --- since it is
physically a distortion of spacetime (in the manner of ripples on
a pond). The existence of these field undulations makes use of the
curved spacetime conceptualization of GR, but at
\em{non}\rm-special-relativistic speeds.

From the above, it appears that the understanding of the links
between GR and energy (and gravitation) may be accepted as
incomplete.
\subsection{Gravitational field oscillations and general relativity}
Sections~\ref{Sect:Eq} and~\ref{Sect:GandE} sought to show that
there appears to be nothing in the foundations of general
relativity, itself, that distinctly forbids a new
(gravito-quantum) mechanism from generating oscillatory
fluctuations upon a pre-existing gravitational
field\footnote{Certainly, the principle of special relativity is
in need of some attention, and the spirit of GR seems to be under
threat, but the provisional acceptance of these type-2
gravitational waves is not unreasonable.}. When a moon rotates
around a planet a similar oscillatory field effect, although
non-constant with radial distance, is experienced by a point mass
--- in the not too near vicinity.

The generation of the waves that affect Pioneer spacecraft (S/C)
are quite distinct from three body effects and general
relativity's gravitational waves. They may be thought of as type-2
gravitational waves if one prefers. Indeed, the existence of
\mbox{type-1} gravitational waves indirectly supports the
existence of another kind of gravitational wave\footnote{A further
source of non-uniformity for a gravitational field is tidal
effects. Such effects ensure lunar spin-orbit coupling around the
moon's host planet.}.
\subsection{Cut-off mass, convolution, and wave energy dispersion}
A sharp (all or nothing) cut-off\footnote{Much in the manner of
the Photoelectric Effect.} exists for the effect of an individual
acceleration wave upon masses in motion.

Inertia involves a body's ability to resist a change in motion.
With the hypothesis being presented, it is also necessary to see
inertia as associated with a wave field's ability to `instill' an
oscillatory variation in the motion of a body.

This situation may be aligned to the mathematical technique of
convolution, which determines a system's output [oscillatory
behaviour: yes or no] given an input signal [the acceleration
wave] and the system impulse response [some function of the
inertial mass of a body]\footnote{Paraphrasing:
\textsf{http://www.see.ed.ac.uk/$^{\sim}$mjj/dspDemos\\/EE4/tutConv.html}
(square brackets content excluded).}.

Although (acceleration) wave amplitudes remain constant, the
(inertial) mass associated with this new kind of wave reduces in
proportion to the volume the dispersing wave
encompasses\footnote{Once again deference to a fuller elucidation
of the model is necessary.}. Thereby, once a type-2 gravitational
wave is generated (or established), conservation of energy is
obeyed\footnote{It is only the \em{generation} \rm of the
acceleration-waves that appears to disobey (expected) conservation
of energy.}.
%-------------------------------------------------------------------------------------------
\section{Amplitudes associated with acceleration waves on the gravitational field}
This section begins the process of quantitative support for the
hypothesis. Two results are established. Results pertaining to the
relationships between the amplitudes of sinusoidal: acceleration,
speed and range variations of a body are outlined. Since the wave
amplitudes (via Lorentz invariance) are necessarily constant, the
mathematics that follows is greatly simplified. Note that the
solar system barycenter acts as the (quasi-)`global' (i.e. solar
system) inertial frame's reference point.
\subsection{Amplitudes of the cyclic variations}
\label{Sect:Amp} The integral of a constant acceleration (of unit
amplitude) over a time $\frac{\pi}{2}$ is simply a velocity of
magnitude $\frac{\pi}{2}$. Now
$\int_{0}^{\frac{\pi}{2}}sin\theta\, \rm{d}\theta=1$ is a velocity
amplitude related to a unit amplitude sinusoidal acceleration
acting over $\frac{\pi}{2}$ (i.e. a quarter of a wavelength). The
period of the wavelength may be either $2\pi$ or $\triangle t$.
Closely related to the maximum sinusoidal acceleration (wave)
amplitude $\triangle a$ and a time of \( \triangle t/4 \) is the
maximum amplitude of a similarly sinusoidal velocity so that:
\begin{displaymath}
\triangle v=\triangle a\cdot \frac{\triangle t}{4}\cdot
\frac{2}{\pi}=\triangle a\cdot \frac{\triangle
t}{2\pi}=\frac{\triangle a}{\omega}
\end{displaymath}
Notice that we let:\\ $|\triangle \overrightarrow{a}|=\triangle
a$, $|\triangle \overrightarrow{v}|=\triangle v$ and $|\triangle
\overrightarrow{x}|=\triangle x$.

In short, there is a direction and time independent relationship
between the magnitudes of velocity and acceleration wave
amplitudes\footnote{By treating magnitudes only we are effectively
now only discussing spacecraft motions beyond Saturn, with
trajectories assumed orthogonal to the wavefronts. Effects from
the smaller moons orbiting Uranus, Neptune and Pluto are
quantitatively negligible.}. This appears in equation 50 of
Anderson et.\,al.~\cite[p.37\,]{And02} with $A_{0}$ replacing
$\triangle a$.

Similarly, we also have for sinusoidal speed variation and
associated ($\frac{1}{4}$wavelength) sinusoidal range change, the
following relationship of amplitudes:
\begin{displaymath}
\triangle x=\triangle v\cdot \frac{\triangle t}{4}\cdot
\frac{2}{\pi}=\triangle v\cdot \frac{\triangle
t}{2\pi}=\frac{\triangle v}{\omega}
\end{displaymath}
Note that: $\omega_{\rm{diurnal}}\approx7.3\times10^{-5}$~rad/s
and $\omega_{\rm{annual}}\approx2.0\times10^{-7}$~rad/s.
Additionally, by way of Ref.~\cite[pp.8, 15, 37]{And02}, the
(return trip) Doppler frequency shift ($\triangle \nu$) is
determined via:
$$\frac{\triangle \nu}{\nu}=\frac{2}{c}\frac{dl}{dt}$$ with the
S-band downlink frequency ($\nu$) being $\sim$2.29 GHz. Finally,
for S-band Doppler: 1 Hz corresponds to 65
mm/s~\cite[p.18]{And02}, or more pragmatically, 10 mHz corresponds
to 0.65 mm/s.
\subsection{Further comments} Observe that if undulations in the gravitational
field are the cause of fluctuations in spacecraft (S/C) speed then
(assuming pure radial motion for the S/C and a position well
beyond Saturn): $\triangle \overrightarrow{a}=-\omega\triangle
\overrightarrow{v}$ and $\triangle
\overrightarrow{v}=\omega\triangle \overrightarrow{x}$.

Note that direction cosines are rampant in this new approach
involving spherical waves emanating from a moon. Small corrections
in the form of direction cosines, for the trajectories of bodies
nearer the center of the solar system, are (quantitatively)
neglected in this write up.

The overall write up is both quantitatively idealized and
theoretically lacking full substantiation, in order to
comprehensively establish the physical validity of the
acceleration-wave hypothesis.
\subsection{A set of annual reference values} \label{Sect:Ref}
The results of Section~\ref{Sect:Amp} applied to an \em{annual}
\rm residual allow a set of reference values to be established.
For $\triangle v=0.2$~mm/s at $\triangle \nu\approx 3.1$~mHz (a
Pioneer S/C S-band Doppler frequency change), the approximate
values of range and acceleration amplitude are: $\triangle x=1$~km
and $\triangle a=0.4\times10^{-8}\rm{cm/s^{2}}$ respectively.
These values are all linearly scalable for different magnitudes of
the four physical `quantities' involved ($\triangle x, \triangle
v, \triangle \nu,\, \mbox{\rm{and}}\, \triangle a$).
%------------------------------------------------------------------------------------------------------------------
\section{Interpreting the Pioneer 10 diurnal residual}
This section shall begin to examine the claim of Anderson et. al.
that: ``[the] annual and diurnal terms are very likely different
manifestations of the same modelling problem~\cite[p.38]{And02}."
\subsection{Introductory remarks}
There are three causes of residuals arising from a method of least
squares analysis. These are: observational error, approximation of
parameters, and model or theory inadequacy~\cite[Ch.8]{Bro61}.
Doppler data cannot clearly discern whether an oscillatory
residual is Earth or spacecraft based. Only the relative
acceleration, motion, or (line segment) distance between the two
bodies is determined by the measurements. Assuming the
observations are reliable, either an oscillatory sinusoidal
residual is a result of Earth based parameter error(s), or failing
that, it is an unlikely \em{real} \rm spacecraft motion and hence
beyond current gravitational theorization.
\subsection{The magnitude of the diurnal residual}\label{Sect:Dmag}
The diurnal residual's interpretation is crucial to understanding
the annual residual. The noise of the diurnal residual is greatest
around solar conjunction, but at solar opposition, near a minimum
in the solar cycle\footnote{There was a broad minimum in the solar
cycle around May 1996.}, exceptionally good data is available:
see~\cite[Fig.18, p.38\,]{And02}.

Markwardt~\cite{Mark02} gives a figure of 10mHz (i.e. 0.65mm/s)
for the amplitude (on average) of the diurnal residuals.
Ref.~\cite[Fig.18, p.38\,]{And02} implies a Nov/Dec 1996 solar
opposition amplitude of $0.1376$~mm/s that may be obtained from:
$$\triangle a=\omega \triangle v$$ and the values of
$\omega_{d.t.}=7.2722\times10^{-5}\rm rad/s$ and
$a_{d.t.}=(100.1\pm7.9)\times10^{-10}\rm{m/s}$ given by
Ref.~\cite[p.38]{And02}. This gives, via $\triangle v=\omega
\triangle x$, a cyclic diurnal position offset of $\sim\pm1.9\rm
\, metres$ (i.e. $\triangle x\approx1.9\,\rm{m}$). As an angular
offset at the Earth's surface this equates to:
$$\theta=\tan^{-1}\frac{1.89}{6378\times10^{3}}\approx1.70\times10^{-5}\rm{deg.}\approx60\rm{mas}$$
where `mas' is milliarcseconds. This being a small and
non-problematic distance or angle, that is less than 0.1\% of
maximum DE\,405 ephemeris error in the Earth's orbit of
$2$~kilometres \cite{Ems04}, and about $16$~times 1997 Earth
orientation polar motion (root-mean-square) calibration accuracy
of $\sim$12 cm~\cite{EOP97}.
\subsection{Diurnal residual parameter groups}
Three main groups of factors or parameters affect the diurnal
residual. For a diurnal effect only short term parameters and
location aspects are included in the first two groups.
\begin{itemize}
\item Earth orientation parameters (EOP) errors.\\This concerns
celestial polar motion offsets, and variability of the earth's
rate of spin. Let us also include here: antenna location errors
for NASA's Deep Space Network (DSN). \item Planetary ephemeris
errors.\\These concern the location of the earth in its orbit
(relative to the solar system barycenter). \item Error via
miscellaneous effects.\\Including: ocean tides, weather, and
variable atmosphere effects; troposphere and ionosphere effects
(causing spectral broadening of the carrier wave frequency); and
interplanetary scintillation (i.e. plasma-based fluctuations).
\end{itemize}
All of this information, and much more, goes into either JPL's
Orbital Determination Program (ODP) or the Aerospace Corporation's
Compact High Accuracy Satellite Motion Program~(CHASMP). The
presence of errors or inaccuracies results in a residual, or what
may be termed a ``spacecraft motion offset". Anderson et. al. have
found that \em{individually} \rm the first two sources of error
cannot be (solely) responsible for the
residual~\cite[p.36]{And02}.

The diurnal residual of~\cite[Fig.18]{And02} may be said to
contain two aspects. Firstly, a stochastic (or random) aspect and
secondly, a cyclic aspect. In the author's opinion, the
miscellaneous effects (discussed above), \em{and} \rm Earth spin
rate variability, predominantly produce either: very small effects
or random shifts in the residual, and hence they may be neglected
from an account of the \em{cyclic} \rm residual indicated in
Fig.18. It is the cyclic diurnal signature that shall now concern
us, as it most closely relates to the annual residual's
interpretation.

Note that the period of the diurnal cycle is not a concern. A
diurnal residual should exist in the Doppler data. The Earth
\em{is} \rm a ``wobbly platform".
\subsection{An interpretation of the cyclic diurnal residual's amplitude}
It appears that only a combination of EOP and Ephemeris error
\em{together,} \rm in the ODP or CHASMP, can (primarily) produce
the cyclic diurnal amplitude error observed at
opposition\footnote{This conclusion is based upon email
correspondence with E. Myles Standish of JPL regarding the diurnal
residual. Note that Myles, whilst having no objection to this
interpretation of the cyclic residual, emphasized the need to not
overlook the other effects mentioned. Also see footnote 125 of
Ref.\,\cite[p.49]{And02}.}.

This interpretation is analogous to how celestial pole offsets
arise. Celestial pole offsets are required because the model (of
the Earth's orientation) that \em{combines} \rm precession and
nutation relies on fixed parameters for the Earth's shape
(geodesy) and internal structure, but since these are not fixed
the offsets necessarily arise. Similarly, combining EOPs and the
ephemeris in an orbital determination program is seen to produce a
(pure) cyclic diurnal residual. This being at 0.1\% of the level
of the maximum error in the DE405 ephemeris (i.e. 2 km).

This seems to indirectly agree with errors in EOPs changing the
value of $a_{p}$ only in the 4th digit \cite[p.36]{And02}. The
diurnal residuals indicate EOPs changing the S/C position location
(only) in the 4th digit of the (Earth position) ephemeris error.
\subsection{In summary}
It appears that a combination of many parameters produces the
diurnal residual, but a combination of two parameter `groups' (EOP
and ephemeris errors) dominates the production of the cyclic
aspect of the diurnal residual --- about a mean value.
%----------------------------------------------------------------------------------
\section{Examining the Pioneer 10 annual residual (1987-1998)} \label{Sect:A1}
An understanding of the diurnal residual allows us to now closely
examine the cyclic `annual' residual. The amplitude of the diurnal
cyclic residual was shown to be primarily a combination of errors
in two parameter groups. Anderson et. al. claim the annual
residual is probably due to a similar combination of parameter
errors. These being: Earth based location and (long-term) polar
orientation errors, and errors in the navigation program's
`determination' of spacecraft (S/C) orbital inclination to the
reference frame being employed.

Note that when averages over a full day or a number of days are
used, the diurnal (and short-term) parameter errors, disappear
from the data. Thus, the annual residual's amplitude is
independent of the amplitude of the cyclic diurnal residual.
\subsection{Orbital inclination errors and the annual residual}
Jet Propulsion Laboratories (JPL), The Aerospace Corporation, and
Craig Markwardt all used DE405 in their `best' analyses.
Interestingly, beginning with DE400 (development ephemeris 400),
both the Earth orientation parameters and the Earth's ephemeris
are aligned to the ICRF (International Celestial Reference Frame).
Error in the orbital inclination of a spacecraft, via an orbital
determination program, is thus more closely related to EOP error
than previously.

A combined outer solar system spacecraft inclination angle error
with Earth polar orientation angle error, is seen to interact with
Earth location (ephemeris) error to produce the annual residual.
This account of the cyclic annual residual, like the cyclic
diurnal residual, is based on two primary parameter groups. This
account although not specifically stated, is alluded to in
Ref.~\cite[p.23, pp.36-38\,]{And02}. It appears a reasonable,
although sketchy, explanation. This scenario is now further
investigated.
\subsection{Quantifying the error associated with orbital
inclination}\label{Sect:Incl} Standish~\cite[p.1166]{Ems04} by way
of the recent (2003) ephemeris DE\,409 has been able to quantify
errors in earlier ephemerides. For all the outer planets
(short-term) inclination errors for both geocentric right
ascension and declination are $0^{"}\!\!.1$ (arc-seconds) for
DE\,200, and $0^{"}\!\!.05$  for DE\,405.

Since the Orbital Determination Programs (ODPs) `produce' the
inclination error, it is best, in this case, to use the error from
DE200 which covered 1979 to 1997. In the case of a predominantly
\em{radial} \rm motion based variation in the Doppler, the effects
of S/C inclination error may be determined\footnote{This, and what
follows, is related to the discussion on p.74 of an article by W.
G. Melbourne~\cite{Melb76} on "space navigation", where range
error is related to an orthogonal distance (and hence angular)
error.}. Assuming radial motion for the distant Pioneer S/C, a
(plus or minus) $0^{"}\!\!.1=$ inclination error for a spacecraft
at 55 AU (Astronomical Unit) implies an uncertainty in position,
orthogonal to the spacecraft's trajectory, of:
$$z=(55)(150\times10^{6})\tan (0^{"}\!\!.1)\approx4,\!000\rm \,km$$
noting that $\tan (0^{"}\!\!.1)\approx4.85\times10^{-7}$. This
orthogonal uncertainty may then be related to a line of sight
uncertainty by a \em{similar} \rm triangle, such that:
$$\triangle x=4\times10^{6}\tan (0^{"}\!\!.1)\approx2\rm \, m$$
where $\triangle x$ is the magnitude of the range variation. Thus,
for $0^{"}\!\!.1$: $z\approx4,000 \rm \,km$ and $\triangle
x\approx2 \rm \, m$; and for $1^{"}$: $z\approx40,000 \rm \,km$
and $\triangle x\approx0.2 \rm \, km$ or $200 \rm \,m$, whereas
for $5^{"}$: $z\approx200,000 \rm \,km$ and $\triangle x\approx5
\rm \, km$.

Thus, on their own the S/C inclination errors arising from DE\,200
fail to account for the `annual' residual, by a long way. At its
\em{minimum} \rm the `annual' range variation amplitude
($\triangle x$) of Pioneer~10 is approximately 500 meters via
$\triangle v\approx0.1\rm{mm/s}$ (see Section~\ref{Sect:Ref} and
Ref.~\cite[p.38]{And02}). Total orientation errors in excess of
1.5 arc-second appear to be required, i.e. above 15 times the
reported orientation error of DE\,200, and 30 times greater than
the DE405 orientation errors of the outer planets.
\subsection{Remarks on combining orbital orientation and ephemeris
errors}\label{Sect:Orient} It should be noted that errors in the
Earth's orbital position (including heliocentric radius error)
will have a minimal cyclic impact upon the errors discussed in
Section~\ref{Sect:Incl}. Since the Earth and Pioneer 10 both lie
very near the plane of the ecliptic, only errors in the Earth's
orbital position \em{far away} \rm from Sun-Earth-Pioneer 10
conjunction or opposition will yield a (line-of-sight) Doppler
range residual. This residual would show narrow peaked maximum or
minimum amplitudes, rather than the smooth `sinusoidal' wave
observed~\cite[Fig.\,1B]{Tury99}.

Finally, with the earlier DE\,200, the ephemerides were oriented
onto their own inherent Earth's mean equator and dynamical
equinox. Thus, the additional orientation error related to the
Earth's annual orbital motion may slightly increase the error, but
this increase would need to be larger than the error in the outer
planets and this is unrealistic.

Subsequently, evidence to the contrary of the Ref.~\cite{And02}
stance on the annual residual is worth considering (this is
pursued in Section~\ref{Sect:A2}). Additionally, let us note that
Markwardt~\cite[p.11]{Mark02} believed: ``\ldots the source [of
the annual residual] was ultimately undetermined."
\subsection{A stance on the `annual' cyclic residual}
The annual spacecraft motion offset, and hence the position
offset, obtained from Doppler tracking observations --- appears to
be neither: a purely orbital inclination effect, nor is it
feasibly a combination of this with either: Earth orbit
orientation error, or Earth position error.

The reason for the `annual' residual appears to be restricted to a
choice between: a mixture of Earth position (ephemeris), Earth
orientation, and spacecraft inclination errors; or alternatively,
an inadequate model of the spacecraft's motion. Concerns regarding
the validity of an account based on parameter errors were raised
in Sections~\ref{Sect:Incl} and~\ref{Sect:Orient}. Indeed, the
``orbital determination programs" need to be of a high quality (in
the first place) to unambiguously obtain the Pioneer anomaly.
%-----------------------------------------------------------------------------------------------------------------
\section{Reinterpreting the Pioneer spacecraft `annual' residuals}
\label{Sect:A2} Section~\ref{Sect:A1} found that spacecraft
orbital inclination error appears insufficient on its own, or
together with Earth ephemeris  and orientation errors, to account
for the `annual' residual's amplitude measured by Doppler tracking
observations. Subsequently, the residual appears to be due to
either: an unrealized ephemeris error of the Earth's position in
its orbit, or the spacecraft has an unmodelled (i.e. real) annual
longitudinal oscillatory motion.

Further concerns may be raised regarding firstly, the cyclic
amplitude of an Earth based explanation of the `annual' residual
of the anomalous Pioneer `acceleration'; and secondly, a new
concern regarding the period of this residual is raised. Beginning
with the latter, these concerns are now addressed, and linkages to
the acceleration-wave hypothesis are drawn upon and incorporated
in the discussion.
\subsection{An alternative approach to the Pioneer `annual' residuals}
On page 38 of the extraordinarily comprehensive Physical Review D
paper~\cite{And02} discussing the Pioneer anomaly, the angular
velocity is given (in interval III) as
$\omega_{a.t.}=(0.0177\pm0.0001)$ rad/day which equates to
$355\pm2$~days per ($2\pi$) cycle. Fig 2. of Scherer et.
al.~\cite{Sch97} shows the real part of the autocorrelation
function of the later Pioneer 10 data (1987-1995). By averaging,
from the graph the clear maximum and minimum range of values, at
the half and full year, a period of $\sim355$~days (and not
$\sim365$~days) is confirmed. The shape of the real part of the
autocorrelation function indicates a solitary sinusoidal-like
oscillation dominates the spectral aspect of a time series
representing the (long-term) Pioneer Doppler data.
Markwardt~\cite[p.11]{Mark02} refers to the ``$\sim$annual"
residuals, also identifying a mean period that is far enough away
from 365.25 days (i.e. 3\%) to be worthy of signification.

The (heliocentric-based) orbital periods of Jupiter's moon
Callisto and Saturn's Titan are respectively: $16.689018$ and
$15.945421$ days. Remarkably, their periods will resonate every
357.9 days ($m=n+1$ where $n\approx21.445$)\footnote{For planetary
ring systems at least: ``Resonances are strongest when $m=n+1$
(for example 2:1 or 43:42) and weaken rapidly as $m$ and $n$
differ more and more~\cite[p.70]{Jewel02}."}. By making a synodic
correction for the location of the Pioneer 10 spacecraft with
respect to the motions of the host planets of these moons (1992.5
to 1998.5), the period goes to approximately
356.1~days\footnote{In the 6 years of interval III (1992.5 to
1998.5) Jupiter tracks, with respect to Pioneer 10's location and
trajectory, approx. $+11^{o}$ prograde, whereas Saturn's position
remains essentially unchanged. Callisto's prograde progression is
thus~$\sim1.8^{o}$ (relative to Titan) per 357.9 day resonant
cycle. This yields a shortening of the ($360^{o}$) resonance cycle
of $\sim1.8$~days. (See Section~\ref{Sect:Fine} for the data
source used to establish these angles.)}. This cycle has a
significant resonance amplitude, and is the only one freely
visible, involving the (proposed) lunar generated
acceleration-waves. Remember, if output from only five moons
dominates the Pioneer anomaly, then only the Callisto-Titan
resonance is expected to be significantly under-affected by the
orbital resonances of Jupiter's Galilean moons. These resonances
act to minimize the (statistical) variance of $a_{p}$ observations
through time\footnote{A small amplitude, directional cosine based,
11.86 year Jupiter cycle should be present, and possibly evident,
in the long term observational data of the anomalous acceleration
[$a_{p}(t)$]. This is perhaps partially evident in~\cite[Fig.14,
p.24]{And02}.}.
\subsection{Pioneer data and the ephemeris}
Any inconsistency that exists between the Pioneer anomaly's annual
residual amplitude and DE405 ephemeris error in the earth's
orbital location of $\pm$1--2\,km~\cite[p.1171]{Ems04} is removed
if the `annual' residual is deemed real. Markwardt~\cite{Mark02}
observes a 10mHz annual amplitude\footnote{Markwardt finds the rms
residuals of all the `non-extreme' Doppler data to be of order
8mHz --- see his Table II. The annual residuals are thus of the
same order of magnitude as the noise, although the noise
amplitude, depending upon space and atmospheric Doppler
transmission conditions, is quite variable.} which implies a
$3.25$~km range error (see section \ref{Sect:Ref} for reference
values). Turyshev et. al.~\cite{Tury99} quote an amplitude of
\mbox{$1.6 \times10^{-8} \rm{ cm/s^{2}}$} implying a range error
of 4 km\footnote{The first oscillation in 1987 has an amplitude of
$2.5 \times10^{-8} \rm{ cm/s^{2}}$ indicative of a $\pm$ 6.25 km
range variation.}. Only later in Interval III can the annual
residual's magnitude be said to be `within expectations'.

Similarly, the month of data given by Anderson et. al. for the
diurnal residual~\cite[Fig.18, p.38]{And02} covers about $30^{o}$
of the annual anomaly's cyclic period. An amplitude increase of
about 0.1\,mm/s (to a maximum) over the 30 days is evident. If
this implies [via $(1-\cos 30^{o})^{-1}=(0.134)^{-1}\approx7.5$] a
speed sinusoid amplitude of $\sim0.75$~mm/s then this roughly
agrees with Markwardt's value of 10mHz or 0.65mm/s
--- for the amplitude of this prospective two-wave resonance\footnote{This is awkward because it is inconsistent with
Ref.~\cite{And02}'s stated amplitude (0.1053~mm/s) for the
interval III annual sinusoid, or 1.5mHz implying a $\pm$~0.5~km
range sinusoid. See Section~\ref{Sect:Damp} for further
discussion.}.
\subsection{Pioneer data and pulsar timing experiments}
Chandler~\cite[p.108]{Chan96} states that: the annual signature of
the Earth's motion dominates the variations of (long term) pulse
arrival times. Additionally, pulsar timing measurements are
accurate to about $3\times10^{-6}$~seconds or 90 meters, and there
has been no indication from pulsar timing experiments of any
overly large error in the Earth's ephemeris (i.e. orbital location
over time) e.g.~\cite[pp.718-19]{Kaspi94}.

The non-problematic account of a slight change in a pulsar's
location by changing from DE200 to DE405~\cite{Japan02}, where
only a 0.2~mas mismatch is apparent, further indicates an absence
of ephemeris accuracy concerns. A similar comment is made by
Anderson et. al. regarding planetary (and spacecraft) ephemeris
error~\cite[p.36]{And02}.

Note that pulsar timing data lack the accuracy of Doppler
\em{diurnal} \rm residual amplitude measurements, that are
approximately 2~meters (or $\sim0.14\rm{\,mm/s}$) around the 1996
Pioneer-Earth-Sun (solar) opposition. Also note that Doppler data
precision reduces as the period of oscillations becomes longer
(given a \em{fixed} \rm oscillation amplitude of frequency
variation $\triangle \nu$).
\subsection{The Pioneer 10 and 11 `annual' residual amplitudes}
A combination of primarily spacecraft inclination, and also Earth
orientation and ephemeris errors has been proposed as a reason for
Pioneer 11's greater `annual' residual amplitude c.f. Pioneer
10~\cite[p.37]{And02}. Alternatively, this feature may be related
to the inclination of the spacecraft trajectories relative to the
\em{equatorial} \rm planes of Jupiter and Saturn. The planet's
equatorial plane is essentially the plane within which Jupiter's
four Galilean moons and Saturn's Titan orbit. Jupiter's and
Saturn's equatorial planes are tilted at 3.1 and 26.7 degrees
respectively, relative to their orbital planes, with their orbital
planes inclined at 1.3 and 2.5 degrees respectively, relative to
the plane of the ecliptic.

Disregarding the inclinations of the planets' orbits, the maximum
possible Callisto-Titan resonance amplitude, from lunar generated
acceleration waves, will thus be approximately at 15 degrees. The
Pioneer 10 and 11 spacecraft trajectories are inclined at
approximately 3 and 16 degrees respectively, to the plane of the
ecliptic. Thus, Pioneer 11 would be expected to have a greater
`annual' cyclic residual amplitude, and probably a slightly
different period to that measured by Pioneer 10 --- if the
hypothesis being outlined in this write up is viable.

If the lunar based acceleration-wave hypothesis is viable then the
other (primary) orbital resonances of Jupiter's moons will
probably play some role in varying the amplitude of this
(Callisto-Titan) resonance over time. This arises because the
other orbital resonances `complicate' the simple decomposition of
this two-wave resonance out of the overall superposition effect of
all the acceleration waves.
\subsection{The fine details of the `reliable' Pioneer observations}
\label{Sect:Fine} A number of details regarding the Pioneer
spacecraft observations appear to be illuminated by the
``acceleration-wave hypothesis". All that is required is to assume
the time-averaged observations are reliable and slightly more
accurate than generally considered. Wave motion to spacecraft
motion direction cosines are implied in all of the following fine
detail aspects.
\begin{enumerate}
\item{Why the overall values of Pioneer 10 and 11 differ
slightly\footnote{Noting that Pioneer 10 has over 11 years of high
quality data, whereas Pioneer 11's data, with only $3\frac{3}{4}$
years, has had insufficient time to establish a representative
`longer-term' average.}.} \item{Why the `annual' residual's
amplitude is greater for Pioneer 11 c.f. Pioneer 10.} \item{Why
the maximum anomalous acceleration for Pioneer 10 is in early
1998.} \item{Why the Pioneer 11 anomaly increases rapidly post
Saturn encounter~\cite[Fig.7, p.18]{And02}.} \item{Why the
magnitude of the Pioneer 10 anomaly is slightly greater in the
later data, as compared to earlier data. (See
Section~\ref{Sect:Disc}).} \item{(Possibly) Why Pioneer 10 spin
rate decreases whereas Pioneer 11's increases between
manoeuvres\footnote{Naturally gas leaks are a likely cause but
``for the Pioneers there were anomalous spin-rate changes that
could be correlated with changes of the exact values of the short
term $a_{p}$. The correlations between the spin-rate changes and
$a_{p}$ a were good to $0.2\times10^{-8} \rm{cm/s^{2}}$ and
better~\cite[p.4019]{NietoV4}."}.}
\end{enumerate}
Point 4 is also influenced by wave direction. At any point within
Saturn's orbit, waves may arrive from Jupiter's Galilean moons and
Saturn's Titan at obtuse angles to each other. Waves from opposing
directions will act to cancel each other out somewhat.
(Section~\ref{Sect:Ener} shall clarify this assertion.)

Regarding point 3: from the Pioneer 10 spacecraft's perspective,
in early 1998 Jupiter and Saturn are closer in the sky to its
`(reverse) trajectory line' than at any other time. They lie near
the spacecraft's reverse (or negative) trajectory
line\footnote{Essentially this is the view (for a forward looking
`passenger') in an imaginary rear view mirror on the spacecraft.}
and approach a conjunction. Hence direction cosine adjustments are
minimized. Saturn actually crosses the extended (negative)
trajectory vector at, or very near to, the early 1998 maximum. See
~\cite[p.24, Fig.14]{And02} as well as planetary and spacecraft
positions given by way of the ``National Space Science Data
Center" website --- to obtain the solar ecliptic reference frame
coordinates through time\footnote{See
\textsf{http://nssdc.gsfc.nasa.gov/space/helios/planet.html} for
planet coordinates and \textsf{.../helios/heli.html} for Pioneer
spacecraft coordinates. An enlargement of Fig.3 in
Ref.~\cite[p.5]{And02} provides a useful adjunct to this data.}.
\subsection{Damping of the annual residual} \label{Sect:Damp}
Fig.\,1B in Ref.~\cite{Tury99} shows the annual residual, as does
Fig.\,13 in Ref.~\cite[p.23]{And02}. The latter figure is divided
into intervals: I, II, and III. Thus, it is reasonable to also
break Fig.\,1B into three intervals. The annual anomaly is most
impressive in intervals I and II (Jan 1987 to July 1992), with a
single sinusoidal line able to smoothly connect essentially every
point (not quite the line shown). Such is \em{not} \rm the case
with Interval III (July 1992 to July 1998) which also contains
amplitudes that are decidedly smaller than in intervals I and II.
Thus, a fairly clear distinction exists either side of July 1992.
The interval III data is also decidedly less sinusoidal than
interval I and II data. Unfortunately, the qualitative detail of
Pioneer 11's annual anomaly is not discussed in Ref.~\cite{And02}.

A pure two-wave resonance based sinusoidal variation should
\em{not} \rm exhibit monotonic damping in its [$a_{p}(t)$] time
series, but there are a number of mitigating circumstances.
Firstly, the orbital resonances of Callisto with Jupiter's other 3
Galilean moons, which are not rigidly fixed
resonances\footnote{Even though Io, Europa and Ganymede are
implied as being in an exact 4:2:1 resonance, their orbital
periods lag/lead relative to each other over time. This full
lead/lag 3 moon cycle takes on average about 1.2 years (i.e.
$\sim14.4$~months.) This duration is based upon ``Sky and
Telescope" magazine's pictorial presentation of Jupiter's moons
over twelve years (1987 to 1998).}, will be influential upon the
Callisto-Titan resonance. Particularly prominent in the damping
could be the method of data smoothing employed. There may also be
another reason, unknown to the author, for this apparent
discontinuity in data `quality' between intervals II and III.
Curiously, the data in interval III is actually preferred to I, II
and Pioneer 11 data by Anderson et. al.~\cite[p.26]{And02} because
the match between JPL's ODP/\em{Sigma} \rm and The Aerospace
Corporation's CHASMP data is so good. Finally, it is worth
mentioning that it is hard to understand how an annual residual
based upon orbital inclination (angle) error could strongly
\em{diminish} \rm with increasing S/C distance from the Earth ---
even taking into account a reduction in Earth and spacecraft
orientation errors over time.

Nevertheless, even with this rationalization the damping is a
concern for the hypothesis being presented in this write up.
Particularly in interval III, and to a much lesser degree in
intervals I and II. This situation could be clarified by a
re-processing of earlier pre-1987
data~\cite{Tury05}~\cite{Giles04}.

\subsection{`Annual' residual -- final remarks}\label{Sect:Afinal}
Bearing in mind Section~\ref{Sect:Damp} there is still an
assortment of evidence that the `annual' residual cannot be a
result of parameter \mbox{error(s)}, as the diurnal residual
surely is. Subsequently, the $\sim$annual residual appears
indicative of a modelling oversight. This model inadequacy suits
the new hypothesis presented, with a Callisto-Titan acceleration
wave `resonance' causing a real oscillatory motion of the S/C
relative to an `accurately' positioned Earth --- itself being
relative to the solar system barycenter. The single sinusoid-like
variation in the autocorrelation data of $a_{p}$~\cite{Sch97},
implies that \em{only} \rm an $(\sim355\pm2)$~day residual is
measured (in Interval III spanning 1992.5 to 1998.5), implying the
($365\frac{1}{4}$~day) `true' annual residual is too small to be
detected. The small changes between DE200 and DE405 for both: the
Pioneer anomaly, and pulsar timing experiments, add support to
this assertion.

It is unlikely a spectral analysis of a very long $a_{p}$
time-series would show any sign of the lesser `true' cyclic annual
spectral component quantified for inclination error in
Section~\ref{Sect:Incl} at approximately $2$~meters. Following the
diurnal case (Section~\ref{Sect:Dmag}), let us estimate total
annual error at $<40$~meters. This signature absence is due to
Doppler data precision diminishing as the period of oscillations
becomes longer (given a \em{fixed} \rm oscillation amplitude of
frequency variation $\triangle \nu$). For the yearly oscillations,
a $0.01$mm/s amplitude, which is around the Doppler tracking's
best level of accuracy, represents a 50 meter range amplitude
(recalling Section~\ref{Sect:Ref}).

The greater precision of a 21st century mission, specifically
designed to test the Pioneer anomaly~\cite{Tury05}, would be
expected to fully clarify this issue. The inclusion of range
measurements would provide a cross-reference for closely
examining, and unambiguously explaining, this solitary
$\sim$\,annual (S/C based), \em{or} \rm `true' annual (Earth
based) cyclic residual present in the Pioneer 10 and 11 data.
%------------------------------------------------------------------------------------------------------
\section{Gravitational field undulations and energy transfer}
\label{Sect:Ener} In light of sufficient concerns regarding the
current account of the annual residual of the Pioneer anomaly, we
return to pursuing the (acceleration-wave) hypothesis sketched
earlier. This section seeks to show that if the $(355\pm2)$~day
cycle is real and spacecraft based (at $\sim356$~days) --- the
origin and explanation of the Pioneer anomaly is open to a
promising and progressive alternative interpretation.
\subsection{Introduction}It has been mentioned by
Anderson et.\,al.~\cite[p.39]{And02} that a sinusoidal speed
variation cannot contribute to the Pioneer spacecraft anomaly.
This is true for an Earth based and/or spacecraft inclination
based residual, e.g. by way of ephemeris errors and Earth
orientation parameters (EOPs) errors in the orbital determination
program (ODP); but this stance may be challenged if the
oscillatory motion is real --- and applied to `light' bodies with
\em{non-zero mass}\rm\footnote{Note that the overall propagation
speed of electromagnetic radiation is necessarily \em{not} \rm
retarded. It ensues that without inertial mass (by definition) the
kinetic energy of E/M radiation particles cannot be incrementally
`eroded'.} e.g. the Pioneer spacecraft. We entertain and examine
this possibility by way of the interaction between spacecraft
`geodesic' motion, and the kinetic and potential \em{energies} \rm
involved. Recall, a geodesic is the closest thing there is to a
straight line in curved spacetime. When (type-2) undulations exist
on the gravitational field, circumstances are different from those
of a `static' gravitational field, especially as regards geodesic
motion.

With \em{no local restoring `force'} \rm to oppose the effects of
a cyclic gravitational undulation, light bodies in celestial
geodesic motion `go with' the (curved spacetime based) cyclic
undulations; whereas the greater inertia of `heavy' bodies means
they fail to respond. For light bodies, spacecraft speed and
distance will oscillate around equilibrium values. A standard
barycenter \em{systemic\footnote{The term ``systemic" is preferred
to ``systematic".}} \rm inertial frame remains valid. With the
passing `phase' of an acceleration wave at the spacecraft, the
local gravitational acceleration field strength varies cyclically,
relative to the system's barycenter inertial frame (and associated
`inertial time').

Even in the absence of a central gravitational field an
oscillatory gravitational field will `disengage' a spacecraft's
constant speed inertial frame, and its geodesic trajectory will
acquire an additional \em{longitudinal} \rm (speed) fluctuation
--- relative to a some systemic inertial reference frame.
\subsection{Agenda}
The aim of Section~\ref{Sect:Ener} is to show that the (short
term\footnote{Especially 1 day, 5 day, or 10 day averages.})
through-time Pioneer anomaly [$a_{p}(t)$] can alternatively be
seen as resulting from a Fourier-like summation and superposition
of `sinusoidal' waves\footnote{Recall that `sinusoidal' means
`harmonic' i.e. some function built up from sinusoidal and
cosinusoidal waveforms.}, i.e. gravitational field undulations (of
amplitude $\triangle a$) upon the gravitational field. The
remainder of Section~\ref{Sect:Ener} seeks to show, by way of
total energy concerns, that the average Pioneer acceleration
($a_{P}=\overline{a_{p}(t)}$) (over a long time) is described by:
$$a_{P}^{2}=\Sigma(\triangle a)^{2} \quad \mbox{or} \quad a_{P}=\sqrt{\Sigma(\triangle a)^{2}}$$
whereas the through-time the instantaneous Pioneer acceleration
$[a_{p}(t)]$ may be expressed by way of:
$$[a_{p}(t)]^{2}=\Sigma[2\triangle a^{2}\cos^{2}(\omega t+\phi)] \quad
\mbox{or more formally:}$$
$$a_{p}(t)=\{\sum_{i=1}^{n}[\,2\triangle a_{i}^{2}\cos^{2}(\omega_{i} t+\phi_{i})]\}^{\frac{1}{2}}$$
where $\phi_{i}$ is the initial (or epoch) phase of a particular
moon's wave, with orbital angular velocity $\omega_{i}$ and wave
amplitude $\triangle a_{i}$ also moon specific.

The relationship between the energy of a spherical acceleration
wave upon the gravitational field affecting a `point' mass through
time, and the oscillatory response of the moving mass to the wave
shall be our prime concern. Recall that (systemically) $\triangle
\overrightarrow{a}$ and $\triangle \overrightarrow{v}$ are
$180^{0}$ or $\pi^{c}$ out of phase.
\subsection{Rayleigh's energy theorem or Parseval's theorem}
\label{Sect:Ray} Rayleigh's energy theorem, sometimes also known
as Parseval's theorem, is an energy-conservation theorem. It says
that a signal\footnote{The terminology describing Fourier
transforms is biased towards electrical engineering interests.}
contains the same amount of energy regardless of whether that
energy is computed in the space/time domain or in the
Fourier/frequency domain. For a continuous function it is:
\begin{equation} \label{eq:Ray}
\int_{-\infty}^{\infty}f^{2}(t)\,\rm{dt}=\int_{-\infty}^{\infty}|g(f)|^{2}\rm{df}
\end{equation}
Thus, the sum (or integral) of the square of a function is equal
to the sum (or integral) of the square of its Fourier
transform\footnote{Based upon:
\textsf{http://research.opt.indiana.edu/Library/Fou\\rierBook/ch$03$.html},
other internet sources, and Ref.~\cite[pp.119-122]{Brac00}.}.
Bracewell~\cite[p.120]{Brac00} notes that: ``... the theorem is
true if [only] one of the integrals exists."

Rayleigh's theorem, by way of ``the power theorem", is also
applicable to the rate of energy transfer, or power, of a
`signal'~\cite[p.121]{Brac00}. A power based application follows
in Section~\ref{Sect:Single}. The relationship between frequency
line width and Fourier transform amplitude in the energy Rayleigh
identity, gives way to a clear distinction between frequency and
energy in the power form of the identity that follows.

By way of the Rayleigh identity, the signal \em{coexists} \rm in
two domains: a time domain, and a frequency domain. In the
application to follow, the signal shall also have two physical
`faces': a field face, and a moving (light or low mass) body face.
\subsection{Gravitational acceleration waves and spacecraft kinetic energy (single cycle, singular
wave case)}\label{Sect:Single} We make use of the Fourier wave
based Rayleigh `power' theorem identity (equation~\ref{eq:Ray} of
Section~\ref{Sect:Ray}), but apply its frequency and particle
aspect, to the new circumstance of spacecraft (S/C) geodesic
motion \em{relative} \rm to a `systemic' (i.e. whole solar system)
\em{inertial} \rm frame centered at the solar system barycenter.
The S/C's geodesic motion `undergoes' a pure sinusoidal (or
cosinusoidal) change in (gravitational) acceleration\footnote{This
acceleration is so tiny that it would be difficult to measure,
even with a very sensitive accelerometer.} and this \em{coexists}
\rm with both: a sinusoidal change in speed ($\triangle v$)
\em{relative} \rm to an equilibrium mean value and, as we shall
soon see, a change (per wave cycle) in barycentric reference frame
speed ($\delta v$).

Since Einstein's General Relativity did not succeed in making
\em{acceleration} \rm relative, this `real' systemic acceleration
is logically capable of representation in a (solar-)systemic
inertial reference frame.

For a \em{single} \rm cycle of a \em{singular} \rm (or isolated)
physical wave upon the gravitational field of: finite extent,
fixed frequency ($F$), and period ($\triangle t$), it follows from
Rayleigh's theorem that:
\begin{displaymath}
\int_{0}^{\triangle
t}f^{2}(t)\,\rm{d}t=\int_{0}^{F}|g(f)|^{2}\rm{d}f
\end{displaymath}
if $f(t)$ is made an even (cosine) function. Note that
$F=(\triangle t)^{-1}$. Regarding the right hand side of the
equation, the single cycle's energy and power are easily expressed
in MKS units\footnote{Mass, kilogram, second units. The frequency
considered is for a single \em{cycle}\rm, but the use of MKS units
requires a frequency with units of [cycles per second].}. Thus,
$|g(f)|^{2}$ is an energy with units of [Joules per cycle] and $F$
has units of [cycles per second]. Noting $|g(f)|^{2}$ is fixed
(per full cycle), and letting $f(t)=\triangle a\cos(\omega t)$, we
have:
\begin{displaymath}
\int_{0}^{\triangle t}[\triangle a^{2}\cos^{2}(\omega
t)]\rm{d}t=|g(F)|^{2}\int_{0}^{F}\!\rm{df}
\end{displaymath}
Bracewell~\cite{Brac00} reassures us that Fourier transforms of
physically real, yet finite sinusoidal waves, do mathematically
exist. With the tiny physical undulations of the gravitational
field ($\triangle a$) physically affecting a moving mass, we
tentatively let the (Fourier transform) amplitude-squared term
equal a specific kinetic energy so that:
$|g(F)|^{2}=\frac{1}{2}(\delta v_{F})^2$ (a physical specific
kinetic energy). Now, since the integral of a squared sine or
cosine function over one period, is half the amplitude squared
multiplied by the period; upon integration and at a given
frequency, we obtain:
\begin{displaymath}
\frac{1}{2}\triangle a^{2}\triangle t=\frac{1}{2}(\delta
v_{F})^{2}F
\end{displaymath}
For convenience, the $F$ subscript on $\delta v$ is dropped, and
we let $F \rightarrow f$, with $f$ now signifying a \em{given} \rm
fixed wave frequency. Noting that $\frac{1}{2}(\delta v)^{2}$ is
written as $\frac{1}{2}\delta v^{2}$, the identity becomes:
\begin{equation}
\label{eq:KE} \frac{1}{2}\triangle a^{2}\triangle
t=\frac{1}{2}\delta v^{2}f
\end{equation}
\subsection{Discussing a single (or unit) cycle of a singular (isolated) wave}
Recalling the dimensions of the identity are $[L^{2}/\,T^{3}]$
which indicates a rate of specific energy transfer. The term
$\frac{1}{2}\delta v^{2}$ is thus indicative of a \em{specific}
\rm kinetic energy \em{transfer} \rm associated with the
sinusoidal oscillation for a \em{single cycle}\rm.

This specific, or mass independent, energy transfer rate is also
being expressed as an unorthodox integral of squared wave
acceleration (i.e. squared field undulation) over time affecting a
body. Note that, at the moment, we are \em{not directly} \rm
dealing with an equality involving wave amplitudes of two physical
quantities ($\triangle a$ and $\triangle v$), but an equality of
specific \rm{energy} transfer rates at, and of, a body. Note that
from the spacecraft's local reference frame it would be the
barycenter that oscillated relative to the spacecraft.
Additionally, note that \em{special} \rm relativistic effects are
negligible in the solar system, while general relativistic effects
are already incorporated in the comprehensive broader analysis
that establishes the Pioneer anomaly.

The two domains involved for this application of Rayleigh's
(Power) theorem apply to: firstly (in the time domain), wave
energy transfer rate (for one cycle) affecting a light body in the
field; and secondly (in the frequency domain), [to] the rate of
oscillatory kinetic energy expressed (in a cycle) by the body. The
former is an effective (specific) energy transfer rate, and the
latter a (specific) energy `expression' rate. Both are `power
quantities' determined relative to a standard systemic inertial
frame (centered) at the solar system barycenter.

Just what is being physically represented, i.e. a field energy
transfer rate being equal to a kinetic energy transfer rate is yet
to be fully clarified.
\subsection{Investigating potential energy}
With the constant phase of the undulations upon the gravitational
field propagating at the speed of light, the distance covered by
the acceleration wave's (constant) phase in a single period of
Jupiter's moon Io (for example) with an orbital period of
$\sim1.77$~days, is about $300$~AU. With the Pioneer spacecraft
travelling at about $2.5$~AU per \em{year}\rm, spacecraft are
effectively stationary as far as the wave's velocity is concerned.
Thus, the variation in gravitational field strength may be
idealized to simply involve the temporal cyclical variation in
gravitational field strength, i.e. we assume changes in S/C
distance from the barycenter are negligible. We further idealize
the situation by assuming constant spacecraft mass. A varying
(specific) potential energy is then simply determined by way of an
oscillating gravitational field strength \em{at} \rm the S/C or
light body. This is a non-central force (or non-Schwarzschild
geometry) based potential energy.

Unlike a central field where (specific) potential energy is
location dependent and proportional to gravitational field
strength, the specific energy of an acceleration \em{wave} \rm is
dominantly period dependent (i.e. time dependent) and proportional
to the square of its amplitude. Thus, for the wave's energy over a
single oscillation, from equation~\ref{eq:KE} it follows that:
$$e_{\rm{wave}}=\frac{1}{2}\triangle a^{2}\triangle
t^{2}$$ This may be thought of as the specific undulation energy
\em{per cycle} \rm \em{of} \rm the (`surface' undulatory)
gravitational field influencing a moving body.

At an essentially fixed barycenter distance, in addition to the
location based (primarily) \em{static} \rm (or non-oscillatory)
potential energy of a body, we now also recognize a time based
\em{oscillatory} \rm specific potential energy associated with a
wave field undulation
--- upon the Sun dominated gravitational field of the solar
system.
\subsection{Relating wave potential energy to kinetic energy (single wave, singular cycle case)}
We now conceptually examine how this additional energy carried by
acceleration waves upon the gravitational field alters the
expression for kinetic energy of a body in motion. We seek to
confirm the specific kinetic energy associated with a single
oscillation is $\frac{1}{2}\delta v^{2}$.

Recalling $\triangle a=\omega \triangle v$ we observe how the
relative oscillatory motion ($\triangle v$, which is also some
function of $\delta v$, see Section~\ref{Sect:amp}), \em{and} \rm
potential energy fluctuations (some function of $\triangle a$) are
$180^{o}$ or $\pi$ radians out of phase. Just how the expression
for specific K.E. per cycle $\frac{1}{2}(\delta v)^{2}$, which is
a response to specific P.E. fluctuations upon the field, is to be
understood is now addressed.

Total energy considerations are different with type-2 undulations
upon the gravitational field. When a sinusoidal acceleration wave
causes a moving mass to display a sinusoidal (output) motion, it
is necessary that the inertial mass `carrying capacity' associated
with the wave's specific potential energy, is just sufficient to
`excite' the moving object, of given mass, into its sympathetic
oscillatory motion response. Thus for the spacecraft:
$$\rm{(P.E._{\rm{wave}})_{\rm{s/c}}}=\frac{1}{2}m_{\rm{s/c}}\triangle
a^{2}\triangle t^{2}$$ If this is the situation, the two
(inertial) masses are seen as common and equal for the wave's
effect upon the moving mass. Hence, we may discuss either:
specific energy equivalence (as in equation~\ref{eq:KE}) or simply
energy equivalence; with the two modes of equality not being
different when resonance \em{is} \rm occurring \em{below} \rm the
cut-off mass $(m_{c})$
--- of a particular wave at a particular \em{location}\rm. Note that below the cut-off mass ($m_{c}$), the effect is independent of
mass. The cut-off mass itself displays a different type of mass
dependence, so that this $m_{c}$ is a function of the \em{wave's}
\rm origin and evolution in space\footnote{The model employs
$\rm{E_{wave}}=\frac{1}{2}m_{\rm{c}}\triangle a^{2}\triangle
t^{2}$ at an initial reference radius, to represent the
\em{wave's} \rm total energy. $\rm{E_{wave}}$ equals the
`gravito-quantum' excess energy ($\rm{E_{excess}}$) `released' to
the gravitational field for (some) third bodies in systemic
three-body motion
--- under the influence of \em{weak} \rm gravitational fields.}.

Relative to the inertial barycentric reference frame, `light'
bodies (e.g. spacecraft) are in both: translational motion
\em{and} \rm a cyclic (line-of-sight) motion. Since the local
frames are not mechanically-forced, unlike the simple harmonic
motion of a spring, there is a specific force (i.e. an
acceleration) restoring back to, \em{and} \rm driving away from,
the equilibrium position. A body's kinetic energy (K.E.)
`reservoir' must be drawn upon to appease the field's oscillatory
demands. It is hypothesized that: the K.E. given to the Pioneer
spacecraft, post-planetary encounter, is being partially directed
by P.E. oscillations (each and every cycle) into a
`non-productive' oscillatory response motion around an equilibrium
value\footnote{The violation of conservation of energy, by way of
an asymmetrical quantum indeterminacy collectively shared by each
and every atom in a moon, expressed in the guise of a field
undulation, has eventuated in this further (or secondary) new
phenomenon.}. In terms of translational kinetic energy, this
oscillatory expression of K.E. is \em{lost} \rm at the rate of
$\frac{1}{2}m_{s/c}(\delta v)^{2}$ per cycle.
\subsection{Relating motion shortfall to the amplitude of sinusoidal speed}\label{Sect:amp}
Returning to the equation (\ref{eq:KE}) equality at the end of
Section~\ref{Sect:Single}, and substituting now for:
\mbox{$\triangle a=\omega\triangle v$} (from
Section~\ref{Sect:Amp}), $\triangle t=1/f$ and $f=\omega/2\pi$ we
obtain:
\begin{displaymath}
\frac{1}{2}(\omega\triangle
v)^{2}(\frac{2\pi}{\omega})=\frac{1}{2}(\delta
v)^2(\frac{\omega}{2\pi})
\end{displaymath}
which simplifies to give $\triangle v=\delta v/\,2\pi$ or:
\begin{equation}
\label{eq:vel} \delta v=2\pi\triangle v
\end{equation}
Thus, the total loss of (solar-)`systemic'
steady\footnote{`Steady' here means non-oscillatory as compared to
oscillatory (i.e. unsteady). Naturally, the steady motion may be:
linear (radial), rotational (i.e. circular or elliptical),
hyperbolic or parabolic.} speed in a \em{single cycle} \rm
($\delta v$), relative to predicted steady speed, equals $2\pi$
times the sinusoidal amplitude of speed variation ($\triangle v$).
Note that this sinusoidal amplitude in speed is about a
\em{non-constant mean} \rm value because kinetic energy
redistribution acts to slow the spacecraft's steady-translational
speed; so that at the beginning and end of the cycle the two
speeds, relative to the (essentially fixed) barycenter, are
different ($v_{\rm{final}}-v_{\rm{initial}}=-\delta v$ where
$\delta v>0$). Finally, note that $|\delta
\overrightarrow{v}|=\delta v$, and that we treat $\delta
\overrightarrow{v}$ predominantly as simply a scalar loss of
speed.
\subsection{Steady and unsteady energy of motion (singular wave case)}
The total kinetic energy, possessed by the Pioneer spacecraft, can
be seen as equal to a steady kinetic energy component plus an
unsteady kinetic energy component. Note that for multiple waves
from multiple moons, of distinct period and amplitude, there will
be a number of coexisting and superpositioned unsteady component
terms.

Importantly, one needs to appreciate that in the absence of any
sinusoidal speed variations all the kinetic energy of a moving
body is steady (i.e. non-oscillatory). This being the case with
general relativistic orbital mechanics, i.e. how the solar
system's gravitation is currently modelled. For the Pioneer
anomaly (in the far outer solar system) small non-conservative
effects, such as radiation pressure and antenna reaction forces,
may be safely neglected.

The motion of both Pioneer spacecraft are predominately steady,
and idealized to be radially directed\footnote{Their outer solar
system hyperbolic motion is idealized to have an infinite radius
of curvature.}. A sinusoid in S/C speed (of amplitude $\triangle
v$), driven by a \em{single} \rm sinusoid in gravitational field
strength (of amplitude $\triangle a$), will redistribute some of
the total (kinetic) energy of a body from fully steady motion into
an additional unsteady (or spectral) component of specific K.E
($\frac{1}{2}\delta v^{2}$ per cycle). This longitudinal
oscillation around an `average' velocity requires kinetic energy,
and this requirement results in a shortfall in translational speed
of $\delta v$ per oscillation --- \em{relative} \rm to the
\em{predicted} \rm condition of fully `steady' motion. This
additional component of K.E. (longitudinal oscillatory for the
Pioneer spacecraft) is `expressed' each and every cycle. Over a
long time these $\delta v$ increments will sum to significant
levels. Over a \em{single cycle} \rm of a singular (effectively
monochromatic) wave we have:
$$(\triangle\rm{K.E.})_{\rm{steady}}=(\rm{K.E.})_{\rm{unsteady}}=\frac{1}{2}m_{s/c}(\delta
v)^{2}$$

The P.E. oscillations act to continuously `erode' the spacecraft's
steady kinetic energy, thus causing a slowing as compared to
predicted motion. This slowing effectively mimics a
non-conservative force-like drag effect upon (the hitherto
assumed) essentially frictionless motion. Think of it as an
acceleration-wave based drag if you like, that `chips' away a
fixed amount of steady K.E. each and every cycle, to `partner' the
P.E. oscillation. Hence, total kinetic energy, i.e. steady plus
unsteady K.E., is also chipped away. Remember this effect only
applies to `light' bodies below the distance dependent cut-off
mass ($m_{c}$) of each specific wave.
\subsection{Interim Remarks}
Consequently, the identity between $\triangle a$ and $\delta v$
(equation~\ref{eq:KE}) derived from Parseval's theorem
(equation~\ref{eq:Ray}), appears to have a justifiable physical
application, i.e. conceivably explaining the Pioneer anomaly. New
physics, beyond the notion of the problematic extra force
hypothesis\footnote{An additional force, i.e. additional spacetime
curvature, cannot explain the Pioneer anomaly's apparent violation
of the principle of equivalence, since the orbits of planets over
a considerable period of time would indicate the presence of such
a `generally applied' force (i.e. the force influences \em{all}
\rm masses). A similar situation applies to the orbits of
long-period (heavy-body) comets~\cite{Whit03}. Additionally, large
outer solar system asteroids (of diameter $>\sim3$~km) will not
exhibit the Pioneer anomaly~\cite{Page05}. See
Section~\ref{Sect:Early}.}, is arising out of a contemporary
interpretation of established mathematics and physics. It is the
`mechanism' behind the establishment, and the associated
quantification of the wave energies and $m_{c}$ magnitudes, that
remains in need of rigorous scientifically explication.

In this write up the waves' existence is simply being hypothesized
and then scrutinized. Partial aspects of a more detailed
theoretical model have been included only to facilitate this
process.
\subsection{Multiple acceleration waves upon the gravitational field}
We are now in a position to examine the effect of the
superposition of acceleration waves, emanating from a number of
moons, upon a spacecraft. Let us continue to idealize the
situation by letting the S/C be in pure radial motion at a
position far outside the solar system, and by letting the total
observation time (i.e. the total data interval) be very long.

Note that the physical mechanism hypothesized, over say 100 years,
has effectively fixed values of: $\triangle a$, $\triangle v$,
$\delta v$ and $\triangle t=1/f$ for any given moon.

From equation (\ref{eq:vel}) (i.e. $\delta v=2\pi\triangle v$) and
$\omega=2\pi f$ it follows that for a singular wave:
\begin{displaymath}
(\triangle v)\omega=(\delta v)f
\end{displaymath}
where $(\delta v)f$, or if you prefer $(\delta v\cdot f)$, is the
rate of speed shortfall (in MKS units) for a singular wave, over a
single wavelength. Now since $\triangle a=\omega\triangle v$:
\begin{eqnarray*}
   \Sigma(\triangle a)^{2} & = & \Sigma\,(\triangle v\cdot\omega)^{2}\\
                     & = & \Sigma\,(\delta v\cdot f)^{2}\\
                     & = & \Sigma\,(\delta v / \triangle t)^{2}
\end{eqnarray*}
Since we are dealing with energy transfer, we must continue to
work with squared quantities, in this case $\triangle a$. Thus,
$\sqrt{\Sigma(\triangle a)^{2}}$ or $[\Sigma(\triangle
a)^{2}]^{\frac{1}{2}}$ is used to represent the overall sum of all
acceleration amplitudes. This sum equals long-term average
(additional) acceleration $a_{P}$ or $\overline{a_{p}(t)}$. Thus,
in the idealized circumstances discussed, it is feasible to write:
$$a_{P}=\overline{a_{p}(t)}=\sqrt{\Sigma(\triangle a)^{2}}=[\Sigma(\triangle
a)^{2}]^{\frac{1}{2}}$$

Now since Doppler tracking measurements, and a raft of supporting
science, give a value of anomalous speed shortfall ($\triangle
V_{P}$), that over a medium-long period of time ($T_{L}$)
indicates the action of an apparently \em{constant} \rm anomalous
acceleration, we may establish that:
\begin{eqnarray*}
 \lim_{t\rightarrow \rm{large}}[\Sigma(\triangle a)^{2}]^{\frac{1}{2}}
 & = & \lim_{t\rightarrow \rm{large}}[\Sigma\,(\delta v / \triangle t)^{2}]^{\frac{1}{2}}\\
 & = & \frac{\triangle V_{P}}{T_{L}}\\
 & = & a_{P}\\
 & = & \rm{constant}
\end{eqnarray*}
Over extended periods of time (i.e. 11.5 years) we have (on
average) a `headline' constant result so that:
$a_{P}^{2}=\Sigma(\triangle a)^{2}=\rm{constant}$. Whereas in the
short term  the $a_{p}(t)$ measured has a stochastic-like
variation\footnote{Actually, the real lunar-based variation is
deterministic if the wave amplitudes and cut-off masses are well
described, and the planetary positions are accurately known. Even
though the acceleration waves are effectively fixed in amplitude
and frequency (over long-terms), the motion of their host planets
and the motion of a spacecraft (or `light' body), will alter the
effect at the spacecraft through time. Thus, the direction-based
variations in amplitude produce a deterministic time series that
coexists with the effects of stochastic Doppler signal noise.}
about the long term fixed mean value ($a_{P}$). Thus, upon
(hypothetically) removing all signal noise, it is proposed that:
$$[a_{p}(t)]^{2}=\Sigma[2\triangle a^{2}\cos^{2}(\omega t+\phi)]$$
best matches the observational data to the hypothesized physical
situation\footnote{The factor of $2$ is necessary, (tentatively)
in order to make the through-time changes of the sum of the
\em{solitary} \rm waves' individual contributions of $\delta
v_{i}(t)$ --- equal to $[a_{p}(t)]$.}. Note that the positional
variation in the additional (Pioneer) acceleration [$a_{p}(x,t)$]
has not been included, as yet, in this idealized model; where $x$
is distance from the barycenter (i.e. from the systemic inertial
frame's `origin').

For a radially (outward) directed spacecraft, at a given position
and a given time, we may \em{schematically} \rm express the total
(inward) acceleration acting on a `light' body as:
$$a_{\rm{total}}(x,t)=a_{\rm{GR}}(x,t)+a_{p}(x,t)-a_{\rm{rad}}(x,t)+\ldots$$
where $a_{\rm{rad}}$ is (specific) outward time dependent solar
radiation force. Note that both the: total \em{instantaneous} \rm
acceleration and this additional (or Pioneer anomalous)
acceleration are now functions of both location and time.
Additionally, in a three (or more) body \em{system} \rm $a_{GR}$,
or standard gravitational theorization, is also time dependent.
The non-determinism of these circumstances produces a need for
ephemerides.

From the above it is also clear that the additional (or Pioneer
anomalous) acceleration ($a_{p}$) is \em{independent of overall
spacecraft speed\rm.
\subsection{Discussion, further remarks, and relaxing the idealization} \label{Sect:Disc}
This section has sought to illustrate that the Pioneer anomalous
acceleration may (alternatively) be hypothesized as resulting from
the effect of a superposition of `sinusoidal' wave undulations
upon a pre-existing gravitational field. Observations and the
author's wider model imply the undulations are lunar based, with
only some (5 out of the 7 major) moons\footnote{For a variety of
reasons the other prograde spin-orbit coupled moons of the solar
system may be neglected. The main reason is that: $\triangle a$ is
proportional to total lunar~mass.} contributing for kinematical
and geometric reasons. Geometry is involved by way of both
spacetime curvature and quantum mechanical geometric phase.

The long term mean amplitude of all the waves together is fixed
(or constant) but over shorter time periods there is necessarily
variation around the long-term mean acceleration. The statistical
variance of observations of $a_{p}(x,t)$ has been greatly
moderated (i.e reduced) by the orbital resonances of the Galilean
moons of Jupiter.

When spacecraft are in the `mid' (10 AU) to outer solar system the
spherical wavefronts are not approximately orthogonal to the
direction of S/C propagation, and directional cosines play a
significant role\footnote{Within the orbit of Saturn things are
different again, with acceleration waves (of significance) being
able to retard a body's motion at obtuse angles to each other,
i.e. in opposing directions. (Recall Section~\ref{Sect:Fine}.)}.
The anomaly will vary with the positions of Jupiter and Saturn
relative to the spacecraft. The effect goes to a slight
\em{asymptotic} \rm maximum as a spacecraft leaves the solar
system and its heliocentric radius gets very large\footnote{The
author's fuller model yields an asymptotic value of:
$a_{P}=\sqrt{[\Sigma(\triangle
a)^{2}]}\approx9.16\times10^{-8}~\rm{cm/s^{2}}$. Once direction
cosines (both in the plane of the ecliptic and inclined to it) are
incorporated, the value of $a_{p}$ for Pioneer 10 (1987 to 1998)
reduces to $\sim8.8\times10^{-8}~\rm{cm/s^{2}}$. This is
embarrassingly, yet significantly, close to the stated result of
$(8.74\pm1.33)\times10^{-8}~\rm{cm/s^{2}}$ given by Anderson et.
al.~\cite{And02}.} (so that direction cosines between the
direction of the moving wavefronts and a body's trajectory
approach one, i.e. unity). Thus, where $x$ is distance from the
barycenter:
$$a_{P}=\overline{a_{p}(x,t)}\approx\rm{constant \qquad} \mbox{if $x\rightarrow$ large}$$
Large $x$ is $>20$\,AU. Surprisingly, due to direction cosines and
the possibility of obtuse angles inwards of Saturn, we have in
general (especially for \mbox{$x<10$\,AU)}:
$$a_{P}\neq\rm{constant \qquad \qquad} \mbox{for all x}$$
which is supported by early Pioneer 11 observations~\cite[Fig.7,
p.18]{And02}.

A second look at the anomaly in the Galileo spacecraft's
\em{cross} \rm solar system journey out to Jupiter (\mbox{Dec
1992-July 1995}) might clarify this issue. Note that a trajectory
or path-based hypothesis, as compared to a Sun-based effect,
involves distinctly different angles, and hence different
anomalous acceleration magnitudes, at these low AU values --- of 1
to 5 astronomical units.

The author's fuller model indicates that at all times the
spacecraft's mass is many orders of magnitude below the `cut-off'
mass at which point a body no longer resonates with at least one
of the undulations upon the gravitational field\footnote{The
existence and magnitude of the cut-off masses, as a function of
distance, indirectly has its origin or source in the mass
\em{dimension} \rm associated with Planck's constant, and
expressions for maximum indeterminacy in quantum mechanical
situations. (Without a fuller model of acceleration wave
`generation', this qualifying footnote is insufficiently
conceptually supported and therefore it may be ambiguous.)}. This
is touched upon in Section~\ref{Sect:Early}.
%----------------------------------------------------------------------------------------------------------------
\section{Further applications of the hypothesis}
The hypothesized existence of: lunar generated, constant
amplitude, first order fluctuations upon the gravitational field
also potentially affects, in a beneficial way, other anomalous
circumstances in our solar system.
\subsection{The Earth flyby anomaly}
Assuming the reality of this anomaly, the preceding hypothesis has
a promising, easy and direct \em{qualitative} \rm application to
the anomalous \em{increase} \rm in the velocity of spacecraft
associated with some, but not all, Earth
flybys~\cite{Flyby98}~\cite{Eqiv01}.

The kinetic energy of the inbound spacecraft will have tiny
additional (longitudinal) oscillatory components of kinetic energy
that are overlooked\footnote{Numerous pre-encounter trajectory
manoeuvres may possibly interrupt and/or overwhelm this
oscillatory component of kinetic energy.}. At Earth gravity assist
encounter, in the planet's reference frame, total S/C energy
remains constant (i.e. is conserved)~\cite[p.449]{Van03}. Then,
post encounter either: the change in trajectory orientation
relative to Jupiter and Saturn reduces the total of the
oscillatory components, or alternatively (and less likely), any
longitudinal oscillatory component is yet to fully reestablish
itself. Subsequently, an excess (steady) kinetic energy over
predictions will be evident on occasions, and the effect will vary
depending upon the pre- and post-encounter trajectories relative
to the positions of (the moons hosted by) Jupiter and Saturn.

It is very difficult for any model to explain an anomalous
\em{increase} \rm in kinetic energy. Other than to cite
observational errors, the hypothesis presented herein is
conceivably the only reasonable non-systematic physical
explanation possible.
\subsection{Early solar system history}
\label{Sect:Early}There are at least two problems concerning the
\em{outer} \rm solar system's very early history\footnote{After
lunar formation and spin-orbit coupling had been attained, around
their respective `host' planets.} for which a new slowing or
`braking' mechanism, in regions of very weak gravitational
fields\footnote{Note that this excludes planetary ring systems.},
may prove beneficial. Like the (predominantly) radial Pioneer
case, orbital motions (in the outer solar system) are subject to a
predominantly radially directed oscillation, that in this case is
now orthogonal to the trajectory of their motion. As for the
Pioneer anomaly, this represents an unsteady expression of kinetic
energy. The inward acceleration of the Pioneer `radial' motion,
for an orbiting body becomes a \em{deceleration} \rm of magnitude
$8.74\times10^{-8}\,\rm{cm/s^2}$ effect. Over a quarter of a
million years this would produce an approximately $7~\rm{km/s}$
slowdown of orbiting `light' bodies\footnote{For bodies inward of
Uranus the deceleration will be a bit less, or slower, than these
values.}, while leaving `heavy' bodies (and electromagnetic
radiation\footnote{The speed of massless electromagnetic radiation
photons also remains (necessarily) unaffected. The dual-aspect
acceleration-waves on the gravitational field, involving both an
oscillatory acceleration magnitude component \em{and} \rm a scalar
inertial mass component, are restricted to only influencing
(light) bodies of non-zero mass.}) totally unaffected. Note that
Neptune's average orbital speed is 5.43 km/s.

For `big rocks' (below the cut-off mass transition), this new
physical `method' of braking light bodies and hence altering their
motion, is much `quicker' than any conceivable braking by existing
conventional drag effects.

The migrating planets hypothesis that is invoked to explain the
\em{too rapid} \rm formation of the \mbox{ice giants} (Uranus and
Neptune), in their present locations, may also possibly be
overcome. This hypothesis is not without its concerns\footnote{For
example, distant Neptune's orbit, although relatively easily
disturbed, is 2nd only to Venus in its circularity (i.e. low
eccentricity).} and criticisms~\cite{Levi01}~\cite{Boss02}. If all
small primordial bodies, in very weak gravitational fields,
underwent `wave-braking' by the aforementioned mechanism then they
would have spiralled into outer planets or the inner solar system.
Compared to other `braking' mechanisms this happens very quickly
and very early in the solar system's history. Naturally, this
mechanism continues to exert its influence today, with collisions
of bodies generating smaller bodies `continuously'.

There is also an apparent total lack of `small' comets observed in
the solar system less than about 1km in
diameter~\cite[Fig.2]{Minor02}, or $\sim5.2\times10^{11}\rm{kg}$
(assuming a comet density of $1\times10^{3}~\rm{kg/m^{3}}$). This
implies that primordial small comets were not flung out, from the
inner solar system, to the Oort cloud long-ago for some reason.
The author's fuller model produces different cut-off masses
($m_{c}$) for each of the five moons previously mentioned --- that
are considered to dominate the Pioneer anomaly. At 2.5~AU from the
\em{lunar} \rm sources, these $m_{c}$ values range between
$3.4\times10^{12}\rm{kg}$ (Titan) to $4.5\times10^{13}\rm{kg}$
(Callisto). Whereas at 20~AU from the lunar sources, these values
go from $6.6\times10^{9}\rm{kg}$ (Titan) to
$8.9\times10^{10}\rm{kg}$~(Callisto). Observational
evidence~\cite[Figs.1 and 2]{Minor02} implies a depletion in
comets, as compared to expectations, beginning at about 5 km
diameter, or a mass of $\sim6.5\times10^{13}\rm{kg}$. Bearing in
mind that primordial comets (orbiting in the plane of the
ecliptic), exist over a range of heliocentric distances with
somewhat random orbital motions; the cut-off masses derived appear
to qualitatively `match' this situation in a very promising
way\footnote{The author's model has not been based upon this
evidence in any way. The observations were found to support the
model \em{after} \rm it was produced to possibly account for the
Pioneer anomaly's observational evidence.}.

In a similar vein, Kuzmitcheva and Ivanov~\cite{Minor02} conclude:
``The lack of small craters on [the asteroid] Eros is the first
observational evidence of a possible paucity of smaller bodies in
the Main Belt~\cite{Bott02}."  Additionally, there apparently
exists an `abrupt' edge to the \em{present} \rm (and presumably
the long-past) solar system by way of an abrupt outer edge to the
Kuiper belt. This was the exact opposite of what planetary
scientists had envisaged~\cite{Malh01}.

Finally, the evidence for this ongoing braking effect on `light'
bodies (of non-zero mass) appears to be supported by the
contemporary representation of a \em{kink} \rm in the linear
relationship between the cumulative number of Earth crossing
asteroids vs. the diameter of these bodies (see the log-log
diagram: Fig.\,3 of Ref.~\cite{Minor02}). There is one straight
line for big primordial bodies, and the other for `fall-out' from
collisions of these bigger bodies. The kink joining the two
possibly illustrates both: the size level at which bigger
primordial bodies start to become sparse, and (at the other end of
the kink) the upper size limit of impermanent collision debris.
%----------------------------------------------------------------------------------------------------
\section{Concluding discussion}
Observational evidence of the `through time' behaviour of the
Pioneer anomaly $a_{p}(t)$ is open to a new interpretation by way
of hypothesizing a number of tiny (sinusoidal) undulations
($\triangle a$) on the gravitational field; each with its own
fixed amplitude and frequency. \em{Ignoring direction cosines}\rm:
$$a_{p}(t)=\sqrt{\sum_{i=1}^{n}[2\triangle a_{i}^{2}\cos^{2}(\omega_{i} t+\phi_{i})]}$$
and, on average over a long period of time, e.g. 30~years (roughly
the orbital period of Saturn):
$$a_{P}=\overline{a_{p}(t)}=\sqrt{\Sigma(\triangle a)^{2}}$$
The acceleration-wave hypothesis is supported, in this write up,
by partial aspects of a new model that seeks to explain the
observational evidence implying at least three anomalous
behaviours in our solar system. The hypothesis, at this stage, is
not easy to accept, because the new physical model discussed has
not been fully established. This needs to be presented to support
and compliment the predominantly general presentation of a
prospective model given in this article. The scope of the
hypothesis is unlike other `singular solutions' offered to explain
a `constant' Pioneer anomaly. Importantly, in this new approach
the apparent violation of the Equivalence Principle is accepted
and incorporated at a fundamental level in the model.

The Pioneer (acceleration) anomaly essentially becomes an
oversight in \em{predicted} \rm spacecraft motion, by way of
omitting the effect of certain wave-like undulations in the
gravitational field's strength. A shortfall in motion results
which is effectively a new non-conservative (specific force)
\em{`deceleration-wave-drag'} \rm effect. The most impressive
evidence for this effect, that is restricted to bodies whose mass
is non-zero and below a certain cut-off mass
zone\footnote{Remember that for each particular acceleration wave
--- the initial (establishment) cut-off mass, described at a characteristic (geometrically based) reference
distance from its lunar source, are (all) slightly different.}, is
the Pioneer anomaly.

This article has sought to show how an ongoing braking effect,
applicable only to \em{light} \rm bodies has sculptured the
contents of our solar system from its very earliest stages.
Somewhat ironically, it may be conjectured that: were it not for
the acceleration-wave-braking mechanism proposed herein, the
Pioneer (and other) spacecraft may not even have passed safely
through the asteroid belt on their journeys to Jupiter and beyond,
many years ago, because it would remain strewn with collision
fragments of \em{all} \rm sizes.
\section{Acknowledgements}
The author greatly appreciated correspondence with E. Myles
Standish of JPL concerning the diurnal residual. Partial funding
was provided by way of an internal grant from the School of
Physics, University of New South Wales. Special thanks to my
supervisor John Webb, and Charley Lineweaver for always having an
answer to a multitude of questions.
%---------------------------------------------------------------------------------------------------------------------

\end{document}